\documentclass[twocolumn,showpacs,preprintnumbers,amsmath,amssymb]{revtex4}

\usepackage{graphicx}
\usepackage{dcolumn}
\usepackage{bm}

\begin{document}

\preprint{APS/123-QED}

\title{Uncertainty Relations for Accurate Measurements}

\author{Seiji Kosugi}
 \email{kosugi@jc.shukutoku.ac.jp}
\affiliation{%
Department of Social Welfare, Shukutoku Junior College, 6-36-4 Maenocho, Itabashi-ku, Tokyo 174-8631, Japan
}%

\date{\today}

\begin{abstract}
Our investigation of the results of the neutron spin experiment by Ehhart et al. demonstrates that their results cannot be understood in accordance with common sense.
For example, their results obtained with different measurement errors are equal to each other for a same neutron state.
This is because their measurement values have a large systematic error due to adopting an incorrect measurement-value operator, which is the operator that gives the measurement values of the measured observable.
They asserted that their results confirmed the validity of a universally valid uncertainty relation proposed by Ozawa.
Therefore, their results demonstrate unexpectedly that Ozawa's uncertainty relation holds true for incorrect measurements.
In addition, we have proved this fact theoretically.
The measurement-value operator must be defined in order that the systematic error of measurement values predicted by the operator is zero.
Such a measurement is called an accurate measurement.
Two kind of uncertainty relations are derived for accurate measurements of a pre-measurement observable $\hat{x}_0$.
In addition, we have presented an uncertainty relation for accurate measurements of a post-measurement observable $\hat{x}_t$.
Heisenberg's original uncertainty relation for measurement processes should be interpreted as this relation between the resolution, that is, the measurement error of the post-measurement observable $\hat{x}_t$ and the disturbance of an object observable $\hat{y}_0$.
\end{abstract}

\maketitle

\section{Introduction}

Recently Erhart et al. \cite{Nat} reported a neutron spin experiment and concluded that the Heisenberg's original error-disturbance uncertainty relation
\begin{eqnarray}
  \epsilon(x_0) \eta(y_0) \geq \frac{1}{2}|\langle \phi_0 | [\hat{x}_0,\hat{y}_0] | \phi_0 \rangle |
\end{eqnarray}
does not hold in general, where $\epsilon(x_0)$ is a measurement error of an observable $\hat{x}_0$, $\eta(y_0)$ is a disturbance of an object observable $\hat{y}_0$ caused by the $\hat{x}_0$ measurement and $| \phi_0 \rangle$ is an initial state of an object system.
On the contrary, a universally valid uncertainty relation
\begin{eqnarray}
  \epsilon(x_0) \eta(y_0)+\epsilon(x_0)\sigma(y_0) + \sigma(x_0) \eta(y_0) \nonumber \\
\geq \frac{1}{2}|\langle \phi_0 | [\hat{x}_0,\hat{y}_0] | \phi_0 \rangle |
\end{eqnarray}
holds always true, where $\sigma(x_0)$ and $\sigma(y_0)$ are initial standard deviations of the observables $\hat{x}_0$ and $\hat{y}_0$, respectively.
This inequality has been proposed by Ozawa \cite{MOPR,MOAP} who is one of the authors of Ref.\cite{Nat}.
We have a few objections to their conclusions.

\section{Heisenberg's original uncertaity relation}

First, the Heisenberg's original uncertainty relation for measurement process is not relation (1).
We agree with them that inequality (1) is not always true. However, inequality (1) is not the Heisenberg's original error-disturbance uncertainty relation.
As has already been pointed out \cite{SK1}, in the thought experiment of the famous $\gamma$-ray microscope \cite{WHZP},
Heisenberg concluded that the uncertainty relation is valid for the electron motion after the measurement \cite{WHPP}.
W. M. de Muynck also remarks that the inequality does not refer to the past but to the future, that is, to the state of the electron after the measurement \cite{Muyn}.
This fact often remains unnoticed.
Therefore, the measurement error in the uncertainty relation must be one that determines the uncertainty in the electron position after the measurement.

We shall here give a brief description of our measurement model.
This is the same as an indirect measurement model \cite{MOAP} that was used by Ozawa in deriving uncertainty relation (2).
We consider a measurement of observables $\hat{x}_0$ of a microscopic object.
The object interacts with a probe, which is a part of an apparatus in a time interval $(0,t)$.
Let $\hat{U}$ be a unitary operator representing the time evolution of the object and probe systems for this time interval.
Then, the object and probe observables $\hat{x}_t$ and $\hat{X}_t$ after the interaction are given by $\hat{U}^{\dagger} (\hat{x}_0 \otimes \hat{I}) \hat{U}$ and $\hat{U}^{\dagger} (\hat{I} \otimes \hat{X}_0) \hat{U}$, respectively. 
We hereinafter abbreviate the tensor product $\hat{x}_0 \otimes \hat{I}$ as $\hat{x}_0$.
The probe is supposed to be prepared in a fixed state $|\xi_0 \rangle$.
After the interaction, the probe observable $\hat{X}_t$ is measured using another measurement apparatus.
Then, using the measurement value $X$ obtained, measurement values of $\hat{x}_0$ and $\hat{x}_t$ are determined.
It is assumed that the probe observable $\hat{X}_t$ can be precisely measured and the measurement apparatus for measuring $\hat{X}_t$ does not interact with the object system.
Such a measurement is called an indirect measurement model.
Every measurement is statistically equivalent to one of indirect measurement models \cite{MOAP}.

We have derived the following inequality in a research of the standard quantum limit (SQL) for monitoring free-mass position \cite{SK2}:
\begin{eqnarray}
  \epsilon_X(x_t) \geq \sigma_X(x_t),
\end{eqnarray}
where $\epsilon_X(x_t)$ is the resolution, that is, the error in a measurement result for the particle position $\hat{x}_t$ just after the measurement when the readout values of the probe position $\hat{X}_t$ is $X$, and the quantity $\sigma_X(x_t)$ is the standard deviation of the particle position $\hat{x}_t$ when the readout is $X$.
Thus, it is the resolution $\epsilon(x_t)$, not the precision $\epsilon(x_0)$ that determines the position uncertainty of the particle after the measurement.
Therefore, the Heisenberg's original uncertainty relation must be interpreted as a relation between the resolution $\epsilon(x_t)$ and the disturbance $\eta(y_0)$:
\begin{eqnarray}
  \epsilon(x_t) \eta(y_0) \geq \frac{1}{2}|\langle \phi_0,\xi_0 | [\hat{x}_t,\hat{y}_t] | \phi_0,\xi_0 \rangle |,
\end{eqnarray}
where we represent the tensor product $|\phi_0\rangle \otimes |\xi_0 \rangle$ as $|\phi_0, \xi_0 \rangle$.
This relation holds always true as will be shown later.

Heisenberg did not distinguish between the measurement errors $\epsilon(x_0)$ and $\epsilon(x_t)$.
When the projection postulate \cite{VNMF,Muyn} of von Neumann holds, inequality (1) can be regarded as the Heisenberg's uncertainty relation especially in the case where the observables $\hat{x}_0$ and $\hat{y}_0$ are the position and momentum operators, respectively, because the measurement errors $\epsilon(x_0)$ and $\epsilon(x_t)$ coincide with each other in this case.
It is obvious in modern quantum theory, however, that the projection postulate does not hold true in general \cite{Muyn}.
In the case of $\epsilon(x_t) \ne \epsilon(x_0)$, inequality (4) must be regarded as the Heisenberg's original uncertainty relation.

\section{Validity of universally valid uncertainty relation}

Secondly, we shall examine the validity of universally valid uncertaity relation (2) proposed by Ozawa.
He derived this relation by using the commutator relation $[\hat{X}_t, \hat{y}_t]=0$ and $(\hat{x}_0)_m \equiv \hat{X}_t$, where $(\hat{x}_0)_m$ is an operator that gives the measurement value of $\hat{x}_0$ and $\hat{y}_t=\hat{U}^{\dagger} \hat{y}_0 \hat{U}$.
This is because they are independent variables.
As has been pointed out in the previous paper \cite{SK1}, in our theory the operator $(\hat{x}_0)_m$ is a function of $\hat{X}_t$.
Suppose that the operator $(\hat{x}_0)_m$ gives correct measurement values. Then, the measurement error
\begin{eqnarray}
  \epsilon(x_0) = \langle \phi_0,\eta_0 | \{ (\hat{x}_0)_{\rm m}- \hat{x}_0 \}^2 |
\phi_0,\eta_0 \rangle ^{1/2}
\end{eqnarray}
satisfies uncertainty relation (2).
In addition, relation (2) holds also for a new measurement error $\epsilon ^{\prime} (x_0)$ defined by using a new measurement-value operator $(\hat{x}_0)_{\rm m}^{\prime}=f((\hat{x}_0)_{\rm m})$, wherer $f(x)$ is an arbitrary function.
When $(x_0)_{\rm m}$ is a measurement value for the operator $(\hat{x}_0)_{\rm m}$, a new measurement value is $f((x_0)_{\rm m})$ in this case.
For example, suppose that $f(x)=100x$. Then, this means that measurement values $100(x_0)_{\rm m}$ also satisfy relation (2).
Because, in general, the operator $(\hat{x}_0)_{\rm m}^{\prime}=100(\hat{x}_0)_{\rm m}$ does not give correct measurement values, this indicates that Ozawa's relation (2) holds for incorrect measurement-value operators.
In other words, his relation is too universal.
Based on the above argument, we can conclude that there exist different uncertain relations for correct measurements (see uncertainty relations (12) and (13)).

\section{Examinations of the results obtained by Erhart et al.}

Thirdly, we shall investigate the validity of thier projective measurement.
Instead of exactly measuring the observable $\hat{\sigma}_x$, they actually carried out the projective measurement of the observable $\hat{\sigma}_\phi={\rm cos}\phi\hat{\sigma}_x+{\rm sin}\phi\hat{\sigma}_y$, where $\hat{\sigma}_x$ and $\hat{\sigma}_y$ are the Pauli matrices and $\phi$ is the detuning parameter.
In this way they regarded the measurement values of the observable $\hat{\sigma}_\phi$ as those of the observable $\hat{\sigma}_x$ and defined the measurement error $\epsilon(x_0)$ as 
\begin{eqnarray}
   \epsilon(x_0) = \langle \phi_0|\{ \hat{\sigma}_\phi-\hat{\sigma}_x \}^2 |\phi_0\rangle ^{1/2}.
\end{eqnarray}
Ozawa proved that there exists a projective measurement for any pure measurement process when the ranges of measurement operators $M_m$ are mutually orthogonal \cite{MOJO}.
He did not show, however, that it is the projective measurement of $\hat{\sigma}_\phi$.
In order to assert that their measurement results of neutron spin experiment prove the validity of uncertainty relation (2), they should present concrete examples of $|\xi_0 \rangle $ and $\hat{U}$ in the indirect measurement model.

In fact, our examinations of their measurement results indicate that regarding the measurement results of $\hat{\sigma}_\phi$ as those of $\hat{\sigma}_x$ is irrelevant.
We shall present two examples here.
First, let $p_+(p_-)$ be a probability that the measurement value is $+1(-1)$. From Fig. 3 in Ref.\cite{Nat}, we obtain $p_+=1$ and $p_-=0$, for the case of $|\phi_0 \rangle =|+y \rangle$ and $\phi=90^\circ$, where $|+y \rangle$ is the eigenket of $\hat{\sigma}_y$ with the eigenvalue$=+1$.
The fact of $p_+=1$ indicates that the same measurement value $+1$ is always obtained when the same measurement is repeated many times.
This result is usually judged that the initial neutron state $|\phi_0 \rangle $ is $|+x \rangle$ and the measurement error $\epsilon$ is zero.
The actual neutron state is, however, $|+y \rangle=(|+x \rangle + |-x \rangle)/\sqrt{2}$ and the measurement error is $2\mathrm{sin}90^\circ/2=\sqrt{2}$.
Such a result cannot be understood reasonably.

Moreover, the standard deviation $\sigma(\sigma_x)$ of the measured observable $\hat{\sigma}_x$ is $1$ in this case.
On the contrary, the standard deviation $\sigma(\sigma_\phi)$ of the measurement values of $\hat{\sigma}_\phi$ is 0.
From the commonsense point of view, the measurement with the measurement error $\sqrt{2}$ increases the standard deviation $\sigma(\sigma_\phi)$ as compared with $\sigma(\sigma_x)$ (see Eq. (10)).
The actual result of $\sigma(\sigma_\phi)$, however, decreases from 1 to 0 as the error increases from 0 to $\sqrt{2}$.
Such a measurement can hardly be recognized as a measurement.

It is found from our detailed investigations that such unreasonable results are caused by the fact that a systematic error $\epsilon_s$ is large.
The systematic error is given by 
\begin{eqnarray}
\epsilon_{\rm{s}} &=& |\Delta|, \nonumber \\
\rm{where} \; \Delta &\equiv& \langle \phi_0,\xi_0  | (\hat{x}_0)_{\rm{m}} |\phi_0,\xi_0 \rangle -\langle \phi_0 | \hat{x}_0 | \phi_0 \rangle.
\end{eqnarray}
In the case of $\sigma(x_0)=0$, the total error is given by $\epsilon=\sqrt{\epsilon^2_{\rm{r}}+\epsilon^2_{\rm{s}}}$, where $\epsilon_{\rm{r}}$ is a random error.
The systematic error expresses the accuracy of the measurement system. It refers to the degree of closeness of measurement values of a quantity to its actual value.
The random error expresses the precision of the measurement system. It refers to the degree of closeness of two or more measurement values to each other under same conditions.

Next, we shall examine their measurement results of $|\phi_0 \rangle =|+z \rangle$.
We obtain $p_+=1/2$ and $p_-=1/2$ from Fig.3 in the case of $\phi=0^\circ$, that is, the measurement error $\epsilon =0$.
Very strangely, we obtain quite the same measurement results for $\phi=40^\circ$ and $\phi=90^\circ$ as those for $\phi=0^\circ$.
The measurement error increases as the parameter $\phi$ increases.
Therefore, this fact indicates that the measurement results of the observable $\hat{\sigma}_x$ with different measurement errors are equal to each other for the same neutron state $|+z \rangle$.
It is sensible to consider that the distributions of the experimental values with different measurement errors are different from each other.
However, they are equal to each other.

More detailed investigations show that the measurement results are the sum of those for $|+x \rangle $ and $|-x \rangle $, as $|+z \rangle = (|+x \rangle +|-x \rangle)/\sqrt{2}$.
The measurement results for $|+x \rangle $ give $p_+=\mathrm{cos}^2 (\phi/2)$ and $p_-=\mathrm{sin}^2 (\phi/2)$ with the error parameter $\phi$, and those for $|-x \rangle $ are $p_+=\mathrm{sin}^2 (\phi/2)$ and $p_-=\mathrm{cos}^2 (\phi/2)$.
Thus, we obtain $p_+=1/2$ and $p_-=1/2$ for $| \phi_0 \rangle=|+z \rangle$ independently of the error, which is in agreement with their experimental result mentioned above.
In the measurements for both $|+x \rangle $ and $|-x \rangle $, the random error $\epsilon_r$ is $\mathrm{sin}(\phi/2)$ and the systematic error$\epsilon_s$ is $1-\mathrm{cos}\phi$.
Thus $\epsilon_r$ and $\epsilon_s$ increase with the error parameter $\phi$.
At $\phi=40^\circ$, we obtain $\epsilon_r\simeq 0.34$ and $\epsilon_s \simeq 0.23$. This indicates that the measurement at $\phi=40^\circ$ is not accurate.

In the case of $\phi=0$, the operator $\hat{\sigma}_\phi$ reproduces the difference of average values $\langle +x |\hat{\sigma}_x|+x \rangle-\langle -x |\hat{\sigma}_x|-x \rangle =2$, because $\langle +x |\hat{\sigma}_\phi|+x \rangle-\langle -x |\hat{\sigma}_\phi|-x \rangle =\rm{cos}\phi-(-\rm{cos}\phi)=2\rm{cos}\phi$.
It does not, however, reproduce the average difference at all in the case of $\phi=\pi/2$, since $2\rm{cos}\phi=0$.
Such a result does not occur when the systematic error is $0$ for both $|+x \rangle $ and $|-x \rangle $.
When $\Delta$ has the same value for any $|\phi_0 \rangle$, we can make the systematic error zero by defining a new measurement-value operator as $(\hat{x}_0)_{\rm{m}}-\Delta$.
For the case of $\hat{\sigma}_\phi$, however, the systematic error can not be made zero in this way, because $\Delta$ is $\rm{cos}\phi-1$ for $|\phi_0 \rangle=|+x \rangle$ and $\Delta=1-\rm{cos}\phi$ for $|-x \rangle$.
It is clear from the above arguments that the measurement values of the operator $\hat{\sigma}_\phi$ cannot be regarded as those of the measured observable $\hat{\sigma}_x$.

\section{Condition of a measurement-value operator}

Our investigations in the preceding section show that measurement results that are not accurate can hardly be understood reasonably.
Therefore, it is reasonable to assume that all measurements must be accurate, that is, the systematic errors $\epsilon_s$ for the measurements must be zero.
Then, we obtain from Eq. (7)
\begin{eqnarray}
  \langle \phi_0, \xi_0|(\hat{x}_0)_{\rm{m}} |\phi_0, \xi_0 \rangle=\langle \phi_0|\hat{x}_0|\phi_0 \rangle .
\end{eqnarray}
Because this condition must be satisfied for any initial object state $|\phi_0 \rangle$, we have 
\begin{eqnarray}
  \langle \xi_0|\{ (\hat{x}_0)_{\mathrm{m}}-\hat{x}_0 \} |\xi_0 \rangle _\kappa =0,
\end{eqnarray}
where $\langle \cdots | \cdots \rangle _\kappa$ represents the partial inner product over the state space of the probe.
This is the same condition of an unbiased measurement.
In other words, the accurate measurement must be the unbiased measurement.
Ozawa often discussed the unbiased measurement \cite{MOAP,MOJO}.
It should be noted, however, that we assert here that the measurement-value operator $(\hat{x}_0)_{\mathrm{m}}$ must be defined to satisfy condition (9) of accurate measurements.

When condition (9) is satisfied, the equation
\begin{eqnarray}
  \sigma^2((x_0)_{\rm{m}})=\sigma ^2(x_0)+\epsilon^2 (x_0)
\end{eqnarray}
holds always true.
It is found from the above equation that such unreasonble measurement results as those obtained by Erhart et al. do not appear for accurate measurements.
When the same measurement value is always obtained for repeated measurements, one has $\sigma((x_0)_{\rm{m}})=0$.
In this case, from Eq. (10) one obtains $\sigma(x_0)=0$ and $\epsilon(x_0)=0$.
Therefore, one judges correctly that the initial object state $|\phi_0 \rangle$ is an eigenket of the measured observable $\hat{x}_0$ and the measurement error is zero.
Moreover, Eq. (10) indicates that the standard deviation $\sigma((x_0)_{\rm{m}})$ increases as the error $\epsilon(x_0)$ increases for the measurements of the object system in the same initial state.
Thus, in this case the distributions of the measurement results obtained with different measurement errors are different from each other.

A fundamental problem of how to decide the operator $(\hat{x}_0)_{\mathrm{m}}$ in quantum measurement theory has hardly been questioned.
A conventional method of determining the operator $(\hat{x}_0)_{\mathrm{m}}$ is as follows \cite{MOPR,MOAP,MOJO,MOPRL,MOPL1,MOPL2}: a certain prove observable $\hat{X}_0$ is regarded to be a meter observable and the observable $\hat{X}_t$ after the interaction is assumed to be the measurement-value operator $(\hat{x}_0)_{\mathrm{m}}$.
No one has examined whether the operator $\hat{X}_t$ gives correct measurement values or not.

The operator $\hat{\sigma}_\phi$ satisfies condition (9) only in the case of $\phi=0$.
This is not surprising, because in this case $\hat{\sigma}_\phi$ is equal to $\hat{\sigma}_x$ and the measurement error $\epsilon(x_0)$ is zero.
An accurate measurement does not mean that its measurement error $\epsilon(x_0)$ is zero. It is a measurement whose systematic error $\epsilon_s$ is zero.
Even if a random error of a measurement is large, the measurement is accurate when $\epsilon_s=0$.

\section{Uncertainty relations for accurate measurements}

We shall drive uncertainty relations that hold for accurate measurements.
Because the measurement-value operator $(\hat{x}_0)_{\rm m}$ is a function of $\hat{X}_t$, we have $[ \; (\hat{x}_0)_{\rm m} , \; \hat{y}_t \; ] =0$.
Thus we obtain 
\begin{eqnarray}
[\; (\hat{x}_0)_{\rm m}  , \; \hat{y}_t - \hat{y}_0 \; ] = -[\; (\hat{x}_0)_{\rm m}-\hat{x}_0, \hat{y}_0 \; ]-[\; \hat{x}_0, \hat{y}_0 \; ].  \nonumber
\end{eqnarray}
Using the condition (9) of accurate measurements, we get
\begin{eqnarray}
\langle \xi_0| [\; (\hat{x}_0)_{\rm m} - \hat{x}_0 , \; \hat{y}_0 \; ] |\xi_0 \rangle_\kappa = 0.
\end{eqnarray}
Thus, $ \langle \phi_0, \xi_0| [\; (\hat{x}_0)_{\rm m}, \; \hat{y}_0 \; ] |\phi_0, \xi_0 \rangle = -\langle \phi_0 | [\; \hat{x}_0 , \; \hat{y}_0 \; ] |\phi_0 \rangle $ is obtained.
Therefore, the following uncertainty relation holds true:
\begin{eqnarray}
  \sigma((x_0)_{\rm m}) \eta(y_0) \geq \frac{1}{2}|\langle \phi_0 | [\hat{x}_0,\hat{y}_0] | \phi_0 \rangle |.
\end{eqnarray}

Moreover, the following inequalities hold \cite{MESSIAH}:
\begin{eqnarray}
  \epsilon(x_0) \eta(y_0) \geq \frac{1}{2}|\langle \phi_0, \xi_0 | [(\hat{x}_0)_{\rm m} - \hat{x}_0,\hat{y}_t-\hat{y}_0] | \phi_0, \xi_0 \rangle |, \nonumber \\
  \sigma(x_0) \eta(y_0) \geq \frac{1}{2}|\langle \phi_0, \xi_0 | [\hat{x}_0,\hat{y}_t-\hat{y}_0] | \phi_0, \xi_0 \rangle |. \nonumber
\end{eqnarray}
Then, 
\begin{eqnarray}
  \{ \epsilon(x_0) + \sigma(x_0) \} \eta(y_0) \hspace{50mm} \nonumber \\
\geq \frac{1}{2}| \langle \phi_0, \xi_0 | [(\hat{x}_0)_{\rm m} - \hat{x}_0,\hat{y}_t-\hat{y}_0] | \phi_0, \xi_0 \rangle \nonumber \\
 + \langle \phi_0, \xi_0 | [\hat{x}_0,\hat{y}_t-\hat{y}_0] | \phi_0, \xi_0 \rangle |. \nonumber
\end{eqnarray}
Because as mentioned above, the equality $\langle \phi_0, \xi_0 | [(\hat{x}_0)_{\rm m} ,\hat{y}_t-\hat{y}_0] | \phi_0, \xi \rangle = -\langle \phi_0 | [\hat{x}_0,\hat{y}_0] | \phi_0 \rangle $ holds, we obtain the following uncertainty relation:
\begin{eqnarray}
  \{ \epsilon(x_0)+\sigma(x_0) \} \eta(y_0) \geq \frac{1}{2}|\langle \phi_0 | [\hat{x}_0,\hat{y}_0] | \phi_0 \rangle |.
\end{eqnarray}
Compared with Ozawa's universally valid uncertainty relation (2), this relation does not include the term $\epsilon(x_0)\sigma(y_0)$.
As has been discussed in Sec.I\hspace{-.1em}I\hspace{-.1em}I, suppose that the measurement-value operator $(\hat{x}_0)_{\rm{m}}$ satisfies relations (12) and (13). A new measurement-value operator $f((\hat{x}_0)_{\rm m})$ does not satisfy relations (12) and (13) in general, whereas it satisfies relation (2).
This is because the operator $(\hat{x}_0)_{\rm{m}}$ must satisfy condition (9) of accurate measurements in our theory.

There has been repeated controversy on the problem of whether the SQL exists for repeated measurements of free-mass position, in particular for gravitational-wave detection \cite{BV,Yuen,Caves,MOPRL,Maddox,SK2}.
In the previous paper \cite{SK2}, we have proved that the SQL for monitoring free-mass position holds always true under the following assumption:
\begin{eqnarray}
  \sigma((x_0)_{\rm{m}}) \ge \sigma(x_0).
\end{eqnarray}
The above inequality is readily derived from Eq. (10).
Therefore, the SQL holds always true for all accurate measurements without assumption (14).
Moreover, uncertainty relation (12) is more fundamental than relation (13), because \begin{eqnarray}
  \epsilon(x_0)+\sigma(x_0) \ge \sigma((x_0)_{\rm{m}}).     \nonumber
\end{eqnarray}

Next, we shall consider the measurement of the observable $\hat{x}_t$ after the interaction.
It is necessary to consider this measurement when one measures the moved distance for a free-mass using the results of the two position measurement \cite{SK2}.
In this measurement, one first measures the post-measurement position $\hat{x}_t$.
The Heisenberg's original uncertainty relation is the relation between the measurement error $\epsilon(x_t)$ for the observable $\hat{x}_t$ and the disturbance $\eta(y_0)$.

As in the case of the measurement of $\hat{x}_0$, the condition of accurate measurements of the observable $\hat{x}_t$ is 
\begin{eqnarray}
  \langle \xi_0|\{ (\hat{x}_t)_{\rm{m}}-\hat{x}_t \} |\xi_0 \rangle_\kappa = 0,
\end{eqnarray}
where $(\hat{x}_t)_{\mathrm{m}}$ is a measurement-value operator of $\hat{x}_t$. The measurement value of $\hat{x}_t$  is also determined  using the measurement value of the probe observable $\hat{X}_t$.
Therefore, the operator $(\hat{x}_t)_{\mathrm{m}}$ is a function of $\hat{X}_t$ and then, we have $[\; (\hat{x}_t)_{\rm m} , \; \hat{y}_t \; ] =0$.
Using the relation 
\begin{eqnarray}
[\; (\hat{x}_t)_{\rm m}-\hat{x}_t, \; \hat{y}_t - \hat{y}_0 \; ] = [\; (\hat{x}_t)_{\rm m}  , \hat{y}_t \; ] - [\; \hat{x}_t, \hat{y}_t \; ] \nonumber \\
- [\; (\hat{x}_t)_{\rm m}-\hat{x}_t, \; \hat{y}_0 \; ],   \nonumber
\end{eqnarray}
we obtain uncertainty relation (4).
Therefore, the Heisenberg's original uncertainty relation holds true in general. When the operators $\hat{x}_0$ and $\hat{y}_0$ are the position and momentum operators, respectively, we have $[\; \hat{x}_t, \; \hat{y}_t \; ]=\hat{U}^{\dagger}[\; \hat{x}_0, \; \hat{y}_0 \; ]\hat{U}= {\rm i}\hbar$.
Then, we obtain
\begin{eqnarray}
  \epsilon(x_t) \eta(y_0) \geq \frac{\hbar}{2}.
\end{eqnarray}

Moreover, the uncertainty relation 
\begin{eqnarray}
  \epsilon(x_t) \sigma(y_t) \geq \frac{1}{2}|\langle \phi_0,\xi_0 | [\; \hat{x}_t, \; \hat{y}_t \; ] | \phi_0,\xi_0 \rangle |
\end{eqnarray}
holds, because $[\; (\hat{x}_t)_{\rm m}-\hat{x}_t, \; \hat{y}_t \; ] = [\; (\hat{x}_t)_{\rm m}  , \hat{y}_t \; ] - [\; \hat{x}_t, \hat{y}_t \; ] $.
This relation is always true in the same manner as Ozawa's universally valid uncertainty relation, because it is derived without condition (15) of accurate measurements. 
This relation is the same as Mensky's uncertainty relation induced by measurement \cite{Mensky}.

\section{Concluding remarks}

We have examined the results of the neutron spin experiment by Erhart et al. and demonstrated that their measurement results can hardly be understood reasonably.
This is because they adopted the incorrect measurement-value operator, not because they made a mistake in their experiment.
They unexpectedly demonstrated that universally valid uncertainty relation (2) holds for such wrong measurement results.
We have also given the theoretical demonstration of this fact.
This is the reason why we cannot accept universally valid uncertainty relation (2) derived by Ozawa.

The measurement values obtained with the measurement error $\epsilon(x_0)>0$ generally differ from the true value.
Thus, we encountered a problem of what a correct measuremnt is for this case.
Our solution of this problem is that it is the accurate measurement where the systematic error is zero.
A preliminary discussion has been held about this problem in the previous paper \cite{SK2}.
Suppose that the operators $\hat{x}_t$ and $\hat{X}_t$ after the interaction are functions of the operators $\hat{x}_0$ and $\hat{X}_0$ before the interaction:
\begin{eqnarray}
   \hat{x}_t = f(\hat{x}_0, \hat{X}_0),   \\
   \hat{X}_t = g(\hat{x}_0, \hat{X}_0),
\end{eqnarray}
where $f(x,y)$ and $g(x,y)$ are arbitrary functions.
For simplicity, we shall consider the case where the state $|\phi_0 \rangle$ is an eigenket of the observable $\hat{x}_0$, that is, $\hat{x}_0$ has only one value.
When the probe state $|\xi_0 \rangle$  is also an eigenket of the observable $\hat{X}_0$, that is, it has no fluctuation of $\hat{X}_0$, the observable $\hat{X}_t$ has only one value in Eq. (19).
Then, the measurement error $\epsilon(x_0)$ is zero.
It is found from this argument that the fluctuation of the probe observable $\hat{X}_0$ causes the measurement error.
Because one cannot know and control which component of the probe wave function $\xi_0 (X_0)$ interacts with the object system in a one-time measurement, it is reasonable to define the measurement-value operators $(\hat{x}_0)_{\rm m}$ and $(\hat{x}_t)_{\rm m}$ as the operators $\hat{x}_0$ and $\hat{x}_t$ obtained by replacing $\hat{X}_0$ by $\langle \hat{X}_0 \rangle$ in Eqs. (18) and (19), respectively:
\begin{eqnarray}
   (\hat{x}_t)_{\rm m} &=& f((\hat{x}_0)_{\rm m}, \langle \hat{X}_0 \rangle),   \\
   \hat{X}_t &=& g((\hat{x}_0)_{\rm m}, \langle \hat{X}_0 \rangle).
\end{eqnarray}
These relations lead to conditions (9) and (15) of accurate measurements.
In addition, using them we can justify a linearity assumption proposed in the previous paper \cite{SK1}, when the operators $\hat{x}_0$ and $\hat{X}_0$ are position operators \cite{SK4}.

Although the accurate measurement always satisfies inequality (14), in the measurement results obtained by Erhart et al., there exists the result that does not satisfy inequality (14) as shown in Sec.I\hspace{-.1em}V.
Theoretically, the standard deviation $\sigma((x_0)_{\rm{m}})$ of measurement values obtained with the measurement error $\epsilon(x_0)=0$ is equal to that $\sigma(x_0)$ of the observable $\hat{x}_0$.
From the point of views, however, that the measurement error cannot be actually reduced to zero and one does not know its value in general before the measurement, in obtaining  the value of $\sigma(x_0)$ by the experiment, the lower limit of the standard deviations $\sigma((x_0)_{\rm{m}})$ obtained by a variety of experiments must be regarded as the value of $\sigma(x_0)$.
Then, only the measurements that satisfy inequality (14) can confirm the validity of the uncertain relation derived by Kennard \cite{Kennard} and Robertson \cite{Robertson}
\begin{eqnarray}
   \sigma(x_0) \sigma(y_0) \geq \frac{1}{2}|\langle \phi_0 | [\hat{x}_0,\hat{y}_0] | \phi_0 \rangle |. \nonumber
\end{eqnarray}
If one recognizes the measurement by Erhart et al. as a measurement, then the validity of the above uncertainty relation cannot be demonstrated by the experiment.

We have derived uncertainty relations (12), (13), and (4) for accurate measurements.
In addition, universally valid uncertainty relation (17) for measuring the observable $\hat{x}_t$ is derived, which corresponds to Ozawa's universally valid uncertainty relation (2) for measuring the observable $\hat{x}_0$.
Uncertainty relation (12) has already been derived by Ozawa \cite{MOAP}.
In Ref. (3), however, he regarded the unbiased measurement as one among many kinds of measurements.
Note that we assert here that the measurement-value operators must be decided in order that the measurement values obtained using them satisfy conditions (9) and (15) of the unbiased measurements, that is, the accurate measurements.
This is because the measurement with a large systematic error shows unreasonable results.

One obtains information about the system under investigation by the measurement.
The information obtained, however, must be intuitively understandable from a commonsense point of view.
Our detailed examinations of the experimental results by Erhart et al. indicate that the measurement-value operator must be defined so that the systematic error is zero.
Although the systematic error cannot be reduced to zero due to many unexpected factors in the actual experiment, the measurement-value operators should be defined in the quantum measurement theory so that it is zero.

This research remind us of what Einstein said to Heisenberg who worried at the contradiction between the electron path observed in the Wilson cloud chamber and his matrix mechanics that denies the concept of "path".
Einstein had said: "It is the theory which decides what we can observe." Heisenberg now felt that the solution of the problem lay in this statement \cite{Jammer}.
Even if the theory predicts the result of a certain experiment and the actual experimental result agrees perfectly with its prediction, however, it does not necessarily prove the validity of the theory.
The experimental demonstration of the universally valid uncertain relation in spin measurement is an example of the above statement.

\end{document}